\documentstyle[aps,eqsecnum,preprint]{revtex}
\tighten
\begin{document}
\title {Utility of Galilean Symmetry in 
Light-Front Perturbation Theory: A Nontrivial Example in QCD}

\author{A. Harindranath and Rajen Kundu}
\address{ Saha Institute of Nuclear Physics \\
Sector I, Block AF, Bidhan Nagar,
Calcutta 700064 India}
\date{February 6, 1998}
\maketitle
\begin{abstract}
Investigations have revealed a very complex 
structure for the coefficient functions accompanying the divergences for
individual time ($x^+$) ordered diagrams in light-front perturbation
theory. No guidelines seem to be available to look for possible mistakes in
the structure of these coefficient functions emerging at the end of a long
and tedious calculation, in contrast to covariant field theory. Since,    
in light-front field theory, transverse boost generator is a kinematical
operator  which acts just as the two-dimensional Galilean boost generator in 
non-relativistic dynamics, it may provide some constraint on the resulting
structures. In this work we investigate the utility of
Galilean symmetry beyond tree level in the context of 
coupling constant renormalization in light-front QCD using the two-component
formalism. 
We show that for each $x^+$ ordered diagram separately, underlying
transverse boost symmetry fixes relative signs of terms in  the coefficient
functions accompanying the diverging logarithms. We also summarize 
the results leading to coupling constant renormalization for the most general
kinematics.   
\end{abstract}
\section{Introduction}
   In light-front field theory, at present, many higher order 
calculations need to be
performed using time ($x^+$) ordered perturbative techniques in
 order to overcome several conceptual and practical problems
\cite{wi94,bpp}. 
Investigations have revealed a very complex structure 
for the coefficient functions (accompanying the divergences)
which emerge at the end of notoriously long and tedious calculations
of individual 
time ($x^+$) ordered diagrams.
It appears that
almost no guidance is available to look for possible mistakes in these
structures.
In contrast, in covariant 
field theory, the
structures accompanying the divergences are quite simple.
The complexity of the former is due to the fact that power
counting is different on the light-front \cite{wi94}. In the latter case, 
simplicity of the structure is due to the underlying Lorentz
symmetry (rotational and boost invariance) which can be maintained at every stage
of the calculation.
Since the light-front formalism do possess some kinematical 
symmetries, it is
worthwhile to investigate whether they
can provide some
constraint on the possible structure of coefficient functions 
for individual $x^+$ ordered diagrams. 

Two of the most important kinematic symmetries in light-front field theory  
which
are relevant especially for phenomenological concerns are the longitudinal and
transverse boost symmetries. The longitudinal boost symmetry is a scale
symmetry on the light-front whereas transverse boost symmetry is simply
Galilean symmetry in two dimensions in non-relativistic 
dynamics \cite{gref}. 
The implications of the Galilean symmetry for the structure of the
interaction vertices resulting from the light-front 
Hamiltonian at tree level are known. 
For example,
the symmetry can be utilized \cite{ln} to reduce 
the number of free parameters in an interaction Hamiltonian constructed 
at tree level purely from light-front power counting. 
The implications of this symmetry beyond tree level is not well-understood.

In this work we investigate the question in the context of coupling constant 
renormalization  in light-front QCD using two-component 
formalism \cite{bks,zh93b}, whether and how the Galilean
symmetry manifests itself beyond tree level 
and whether the symmetry can
provide some guidance in understanding the complex structure of coefficient
functions accompanying the divergences. Specifically, we study the
corrections to the canonical quark-gluon vertex arising from quark-gluon and
three-gluon vertices. It turns out that the accompanying structures are
either proportional to the canonical vertex or independent of the total
momentum
momenta and thereby exhibiting transverse boost invariance. 
Incidentally, these 
processes are relevant for the calculation of
asymptotic freedom in light-front QCD and we also present the 
$\beta$-function calculation for the most general kinematics in the
two-component formalism. However, our 
motivation for 
studying
these processes is quite different in the present case. 
We want to stress the role played by Galilean boost symmetry in
ensuring the correctness of the structure of coefficient functions appearing
in the calculation beyond tree level for each $x^+$ ordered diagram
separately. Such nontrivial checks are extremely helpful,
for example, in extending
the calculations of structure functions presented in Ref.(\cite{pp}) 
to the next higher order. For completeness, we
mention that previous calculation \cite{perry} 
of vertex corrections have employed the four-component representation
 of Brodsky and
Lepage \cite{bl} and only the final answers after summing different time
orderings have been presented. A calculation \cite{zh93c} of the vertex
correction using the
two-component representation have studied only the two 
specific cases of 
helicity-flip part of the vertex (proportional to quark mass) and zero momentum
($q^{+,\perp}=0$) limit for the gluon. For the investigation of 
Galilean symmetry of the vertex beyond tree level
we need to study each time-ordered
contribution separately for
arbitrary momenta for the external legs, which we undertake in this work.      

 The plan of this paper is as follows. In section 2, we discuss aspects of
transverse boost symmetry in light-front dynamics. In section 3, explicit
calculations are carried out for the quark-gluon vertex at the first
non-trivial order. In section 4, we confirm the $\beta$-function calculation
presented in Ref.(\cite{zh93c}) for the most general kinematics.
Finally, a summary, discussion of the results, and 
their possible implications are provided in section 5.      
\section{Transverse Boost Symmetry: Canonical Considerations}
In light-front theory, the generators of transverse boost are given by
\begin{eqnarray}
E^i ~= ~M^{+i} ~= ~{1 \over 2} ~\int dx^- d^2 x^\perp ~\Big [ ~x^i ~
\theta^{++} ~ - x^+ ~ \theta^{+i} ~ \Big ]
\end{eqnarray}
where $\theta^{\mu \nu}$ is the symmetric energy-momentum tensor.
The generators $E^i$ leave $x^+=0$ invariant and hence are kinematic
operators. They obey the 
commutation relations 
\begin{eqnarray}
\Big [ ~ E^i, P^j ~ \Big ] ~ = ~ -i \delta^{ij} P^+~,~ \Big [ E^i, P^- \Big
] ~= ~-2 i P^-~,~ \Big [ J^3,E^i \Big ] ~=~ i \epsilon^{ij} E^j~, 
\end{eqnarray}
where $P^+$ and $P^i$ are the longitudinal and transverse momentum operators
respectively,
$P^-$ is the Hamiltonian operator, $J^3$ is the helicity operator, and
$\epsilon^{ij}$ the two dimensional antisymmetric tensor.
Thus the generators $E^i$ act just like Galilean boosts in 
the transverse plane, familiar from non-relativistic dynamics. 

In light-front theory involving fermions and gauge bosons, the interaction
vertices have a nontrivial structure.
Galilean symmetry implies that the interaction vertices in the
theory (in momentum space) are independent of the total transverse momentum
in the problem. Since the issues associated with Galilean invariance are
most transparent in the
two-component representation, it is most convenient to 
use this representation \cite{bks,zh93b} of light-front QCD in
contrast to the more familiar four-component representation \cite{bl}. In
this work we
follow the conventions of \cite{zh93b} except for the commutation
relations of 
fermionic Fock operators which we choose as        
\begin{eqnarray}
\big \{ b(p,\lambda), b^\dagger(p',\lambda')\big \} = 
\big \{ d(p,\lambda), d^\dagger(p',\lambda')\big \} = 
2 (2 \pi)^3 p^+
\delta_{\lambda, \lambda'} \delta^3(p-p').
\end{eqnarray}
Accordingly, in our notation, the two-component fermion field is given by 
\begin{eqnarray}
\xi(x) = \sum_{\lambda}~ \chi_\lambda ~\int {d p^+ d^2 p^\perp \over 2 (2
\pi)^3 \sqrt{p^+}} ~\big [ b(p,\lambda)~e^{-i p.x} ~+~ d^\dagger(p, -\lambda)
~e^{i p.x} \big]. 
\end{eqnarray}

The canonical quark-gluon vertex in our notation is (see Fig. 1)  
\begin{eqnarray}
{\cal V}_1 ~= &&~g ~T^a ~ \sqrt{p_1^+~p_2^+}~ \chi^\dagger_{s_{2}}~
 \Big [ - 2 {q^\perp \over q^+}
+ \sigma^\perp {\sigma^\perp . p_1^\perp \over p_1^+} + {\sigma^\perp
.p_2^\perp \over p_2^+} \sigma^\perp  +im \Big( {1 \over p_1^+} - {1 \over
p_2^+} \Big) \Big ] ~\chi_{s_{1}} ~ .~
(\epsilon^\perp_\lambda)^*. \nonumber  \\ \label{canon}
&&
\end{eqnarray}
Since the mass term (helicity-flip interaction) is irrelevant for the
Galilean invariance, we drop it in the following.
Note that the vertices and energy denominators in
$x^+$ ordered diagrams in the two-component representation 
are manifestly invariant under Galilean boost. 
(See  the appendix for an explicit example at one loop level).
    
\section{One loop calculations}
 
In the {\it massless} limit, the helicity-flip contribution 
vanishes and the canonical vertex has the
structure
\begin{eqnarray}
{\cal V}_1 ~= ~g ~T^a ~ \sqrt{p_1^+~p_2^+}~ \chi^\dagger_{s_{2}}~
 \Big [ - 2 {q^\perp \over q^+}
+ \sigma^\perp {\sigma^\perp . p_1^\perp \over p_1^+} + {\sigma^\perp
.p_2^\perp \over p_2^+} \sigma^\perp   \Big ] ~\chi_{s_{1}} ~ .~
(\epsilon^\perp_\lambda)^*. 
\end{eqnarray}
In this section we consider corrections to this vertex at one loop level
in LF Hamiltonian perturbation theory.
Specifically we consider the corrections arising from quark-gluon
vertex and the three-gluon vertex. Note that the corrections arising from
instantaneous vertices in the theory do not contribute to the divergent
structure of the vertex for zero quark mass at one loop level.

In order to perform the calculations beyond tree level, we need to
regulate the loop momenta. 
How to introduce regulators in light-front theory is, at present, an active 
subject of research \cite{ann}. One may (1) introduce cutoffs on the 
sum of light-front energies (the so-called boost invariant cut-off), or 
(2) choose to cutoff energy
differences at vertices (which emerge naturally in similarity renormalization
perturbation theory) or (3) simply cutoff single particle momenta. 
We employ the third choice for the regulators, namely, $ k_i^+ > \epsilon$,
$ \mu < k_i^\perp < \Lambda$, which is 
simple to implement but obviously violate both longitudinal
and transverse boost invariance.  
Since the vertices and energy denominators are explicitly invariant under
the Galilean boost (see the appendix), the violation of the symmetry can
occur only through the explicit appearance of total transverse momentum 
$P^\perp$ in the limits
of integration. From power counting, the vertex corrections at one loop
level are only logarithmically divergent in the transverse plane. 
Hence we expect the coefficient functions accompanying the logarithms to
still exhibit the symmetry.

Consider the one loop corrections to the vertex involving two
quark-gluon vertices. There are two time-ordering contributions shown in
Figs. 2(a) and 2(b). 
The contribution from Fig. 2(a) is 
\begin{eqnarray}
{\cal V}_{2a} ~=&&~{g^3 \over 2 (2 \pi)^3}~ T^b T^a T^b ~ \sqrt{p_1^+~p_2^+} 
\int_\epsilon^{p_2^+-\epsilon} dk^+ \int d^2 k^\perp 
~ \theta(\Lambda - \mid k^\perp \mid) \nonumber \\
&& ~~~~~~{ 1 \over k_3^+}~ 
{ 1 \over p_1^- - k_1^- - k_3^-} ~{ 1 \over p_1^- - q^- - k^- - k_3^-}
\nonumber \\
&& ~~~~~~ \sum_{\sigma_1 \sigma_2 \lambda_1}
~\chi^\dagger_{s_{2}}~ 
\Big [ -2  {k_3^\perp \over k_3^+} + \sigma^\perp {\sigma^\perp .k^\perp \over
k^+} + {\sigma^\perp . p_2^\perp \over p_2^+} \sigma^\perp \Big ]~ 
\chi_{\sigma_1} ~. ~\epsilon^\perp_{\lambda_{1}} ~ \nonumber \\
&& ~~~~~~~~~\chi^\dagger_{\sigma_1} ~
\Big [ -2  {q^\perp \over q^+} + \sigma^\perp {\sigma^\perp .k_1^\perp \over
k_1^+} + {\sigma^\perp . k^\perp \over k^+} \sigma^\perp \Big ] ~
~\chi_{\sigma_{2}} ~. ~(\epsilon^\perp_{\lambda})^* 
\nonumber \\
&&~~~~~~~~~\chi^\dagger_{\sigma_{2}} 
~\Big [ -2  {k_3^\perp \over k_3^+} + \sigma^\perp {\sigma^\perp .p_1^\perp 
\over p_1^+} + {\sigma^\perp . k_1^\perp \over k_1^+} \sigma^\perp \Big ]~ 
\chi_{s_1} ~. ~(\epsilon^\perp_{\lambda_{1}})^*. \label{eq2a}
\end{eqnarray}
Here $ k_1^{(+,\perp)} = q^{(+,\perp)} + k^{(+,\perp)}$ and $ k_3^{(+,\perp)}
 =
p_2^{(+,\perp)} - k^{(+,\perp)}$. 

After a long and tedious calculation, we arrive at two types of divergent
contributions, one containing  product of logarithms and other containing a
single logarithm. We have taken $\mu$ and $\Lambda$ to be much larger 
than the external momentum 
scales in the problem.
Divergent contributions that contain products of logarithms:
\begin{eqnarray}
{\cal V}_{2a}^{I}~=&&~g ~T^a ~ 
\sqrt{p_1^+~p_2^+} ~\chi^\dagger_{s_{2}}~  
\Big [ - 2 {q^\perp \over q^+} +  \sigma^\perp { \sigma^\perp .p_1^\perp \over
 p_1^+} +  { \sigma^\perp . p_2^\perp \over p_2^+} \sigma^\perp \Big
]~\chi_{s_{1}}~. ~(\epsilon^\perp_\lambda)^*~ \nonumber \\
&&~~~~~~~~ {g^2 \over 8 \pi^2} ~ \Big( -{1 \over 2}~ C_A ~+ ~C_f \Big)~
ln{\Lambda \over \mu} ~  4 ~ln{p_2^+ \over \epsilon}.\label{eq2a1}
\end{eqnarray}
Since the coefficient of the divergent factor is proportional to the
canonical vertex, the transverse boost invariance of the above result is manifest. 
Divergent contributions that contain single logarithm are 
\begin{eqnarray}
{\cal V}_{2a}^{II}~ = && ~g ~ T^a~
\sqrt{p_1^+~p_2^+} ~\chi^\dagger_{s_{2}}
 ~\Big [ 
6 {q^\perp \over q^+} - 6 {p_1^\perp \over p_1^+}
 - {\sigma^\perp . p_2^\perp \over p_1^+}\sigma^\perp
+ {p_2^+ \over p_1^+} {\sigma^\perp . p_1^\perp \over p_1^+} \sigma^\perp 
 \Big]~\chi_{s_{1}}~. ~(\epsilon^\perp_\lambda)^*~ \nonumber \\ 
&& ~~~~~~~~~{ g^2 \over 8 \pi^2}~ \Big ( -{1 \over 2}~ C_A~ + ~C_f \Big )~
 ln{\Lambda \over \mu} \label{eq2a2}.
\end{eqnarray}
In this case the coefficient of the divergent factor involving transverse
momenta is not proportional to the canonical vertex. However, in terms of
the internal momenta (see the appendix), the quantity inside the square
bracket can be rewritten as
\begin{eqnarray}  
\Big [ 
6 {q^\perp \over q^+} - 6 {p_1^\perp \over p_1^+}
 - {\sigma^\perp . p_2^\perp \over p_1^+}\sigma^\perp
+ {p_2^+ \over p_1^+} {\sigma^\perp . p_1^\perp \over p_1^+} \sigma^\perp 
 \Big] \nonumber \\
=-{1 \over P^+} \Big [{ 6 \kappa_1^\perp \over 1-x} + \sigma^\perp
.\kappa_1^\perp \sigma^\perp \Big ],
\end{eqnarray}
which satisfies the constraint from  Galilean invariance.

Contribution from Fig. 2(b) is    
\begin{eqnarray}
{\cal V}_{2b} ~= &&~
(-) ~{g^3 \over 2 (2 \pi)^3} ~ T^b T^a T^b ~ \sqrt{p_1^+~p_2^+}
~\int_\epsilon^{q^+-\epsilon} dk_2^+ ~
\int d^2 k_2^\perp ~ \theta(\Lambda - \mid k^\perp \mid)~ { 1 \over p_1^+ - k_2^+}
~{ 1 \over p_1^- - k_1^- - k_2^-} ~\nonumber \\
&& ~~~~~{ 1 \over p_1^- - k_2^- - k_3^- - p_2^-}~ 
\sum_{\sigma_1 \sigma_2 \lambda_1}~ 
\chi^\dagger_{s_{2}} 
~\Big [ -2  {k_1^\perp \over k_1^+} + \sigma^\perp {\sigma^\perp .k_3^\perp \over
k_3^+} + {\sigma^\perp . p_2^\perp \over p_2^+} \sigma^\perp \Big ] 
~\chi_{\sigma_{1}} ~.~ \epsilon^\perp_{\lambda_{1}} ~ \nonumber \\
&& ~~~~~~~~~\chi^\dagger_{\sigma_1} 
~\Big [ -2  {q^\perp \over q^+} + \sigma^\perp {\sigma^\perp .k_2^\perp \over
k_2^+} + {\sigma^\perp . k_3^\perp \over k_3^+} \sigma^\perp \Big ] 
~\chi_{\sigma_{2}} ~. ~(\epsilon^\perp_{\lambda})^*  \nonumber \\
&& ~~~~~~~~~\chi^\dagger_{\sigma_{2}} ~
\Big [ -2  {k_1^\perp \over k_1^+} + \sigma^\perp {\sigma^\perp .p_1^\perp 
\over p_1^+} + {\sigma^\perp . k_2^\perp \over k_2^+} \sigma^\perp \Big ] 
~\chi_{s_1} ~. ~(\epsilon^\perp_{\lambda_{1}})^*. \label{eq2b}
\end{eqnarray}
Here $ k_1^{(+,\perp)} = p_1^{(+,\perp)} - k_2^{(+,\perp)}$ and 
$ k_3^{(+,\perp)} =
q^{(+,\perp)} - k_2^{(+,\perp)}$. The overall -ve sign arises from the anti
symmetry property of fermionic states. Note that this -ve sign is missing
from Eq. (A8) of Ref.\cite{zh93c}.

As in the previous case, explicit evaluation leads to terms containing
two types of divergences.
Divergent contributions that contain products of logarithms are
\begin{eqnarray}
{\cal V}_{2b}^{I}~ = &&~ g ~ T^a 
~ \sqrt{p_1^+} ~\sqrt{p_2^+} ~\chi^\dagger_{s_{2}}~ 
 \Big [ - 2 {q^\perp \over q^+} +  \sigma^\perp { \sigma^\perp .p_1^\perp 
\over
 p_1^+} +  { \sigma^\perp . p_2^\perp \over p_2^+} \sigma^\perp \Big ]~
\chi_{s_{1}}~. ~(\epsilon^\perp_\lambda)^*~ \nonumber \\
&& ~~~~~~~ {g^2 \over 8\pi^2}~ \Big ( -{1 \over 2}~C_A ~+ ~C_f \Big )~
 ln{\Lambda \over \mu} ~  2 ~ ln{p_1^+ \over p_2^+} . \label{eq2b1}
\end{eqnarray}
Again, the transverse boost invariance of this result is manifest since the
contribution is proportional to the canonical vertex.
Divergent contributions that contain single logarithm  are 
\begin{eqnarray}
{\cal V}_{2b}^{II}~ = &&~(-) ~g ~T^a
~ \sqrt{p_1^+~p_2^+} ~\chi^\dagger_{s_{2}}~  
\Big [ 
 3 \sigma^\perp {
\sigma^\perp.p_1^\perp \over p_1^+} 
+ 3 {\sigma^\perp .p_2^\perp \over p_2^+} \sigma^\perp \nonumber \\
&&- 6 {p_1^\perp \over p_1^+} - {\sigma^\perp . p_2^\perp \over p_1^+}
\sigma^\perp
+ {p_2^+ \over p_1^+} {\sigma^\perp . p_1^\perp \over p_1^+} \sigma^\perp
 \Big]~ 
\chi_{s_{1}}~.~ (\epsilon^\perp_\lambda)^* \nonumber \\
&& { g^2 \over 8 \pi^2} ~ 
\Big ( -{1 \over 2}~C_A ~+~ C_f \Big )~ln{\Lambda \over \mu}. \label{eq2b2}
\end{eqnarray}
The transverse boost symmetry of the terms inside the square bracket is not
manifest but becomes explicit once we express the result in terms of the
internal momenta. Alternatively, 
by subtracting and adding the term  $ - 6 {q^\perp \over q^+}$ to these
terms we can rewrite the terms inside the square bracket as the canonical
term plus the terms contained in the square bracket in eq.(\ref{eq2a2})
which again shows the boost invariance of the result in eq.(\ref{eq2b2}).

Consider, next, the one loop contributions to the quark-gluon vertex
involving one quark-gluon vertex and one three gluon vertex. There are two
time ordering contributions shown in Figs. 3(a) and 3(b).
The contribution from Fig. 3(a) is
\begin{eqnarray}
{\cal V}_{3a}~=&&~ {g^3 \over 2 (2 \pi)^3}~ (-i f^{abc} T^b T^c)
 ~ \sqrt{p_1^+~p_2^+}
~ \int_\epsilon^{p_2^+-\epsilon} dk^+
~ \int d^2 k^\perp ~ \theta(\Lambda - \mid k^\perp \mid)
~{ 1 \over k_1^+} ~{1 \over k_2^+} ~{ 1 \over p_1^- - k_1^- - k^-}
\nonumber \\
&& ~~~~~{ 1 \over
p_1^- - q^- - k_2^- - k^-}~
~ \sum_{\sigma_1, \lambda_{1}, \lambda_{2}} 
~\chi^\dagger_{s_{2}} ~\Big [ -2 { k_2^\perp \over k_2^+} + \sigma^\perp
{\sigma^\perp . k^\perp \over k^+} + {\sigma^\perp. p_2^\perp \over p_2^+}
\sigma^\perp \Big ] ~\chi_{\sigma_1} ~.~ \epsilon^\perp_{\lambda_{2}}~ 
\nonumber \\
&&~~~~~~~~ \chi^\dagger_{\sigma_1}~ \Big [ -2 { k_1^\perp \over k_1^+} + \sigma^\perp
{\sigma^\perp . p_1^\perp \over p_1^+} + {\sigma^\perp. k^\perp \over k^+}
\sigma^\perp \Big ]~ \chi_{s_{1}} ~. ~(\epsilon^\perp_{\lambda_{1}})^* 
\nonumber \\
&&~~~~~~\epsilon^j_{\lambda_{1}} ~(\epsilon^i_{\lambda})^*~
(\epsilon^l_{\lambda_{2}})^* ~\Bigg [ \Big [ (k_1^i + k_2^i) - {q^i \over
q^+} (k_1^+ + k_2^+) \Big ] \delta_{lj}
- \Big [ (k_1^l + q^l) - {k_2^l \over k_2^+} (k_1^+ + q^+) \Big ] \delta_{ij}
\nonumber \\
&&~~~~~~~~+  \Big [ (q^j - k_2^j) - {k_1^j \over k_1^+} (q^+ - k_2^+) \Big ]
\delta_{il} \Bigg ] .  \label{eq3a}
\end{eqnarray}
Here $ k_1^{(+,\perp)} = p_1^{(+,\perp)} - k^{(+,\perp)}$ and 
$ k_2^{(+,\perp)} =
p_2^{(+,\perp)} - k^{(+,\perp)}$.
Divergent contributions that contain products of logarithms are
\begin{eqnarray}
{\cal V}_{3a}^{I}~=&& ~g ~T^a~  ~\sqrt{p_1^+~p_2^+} ~\chi^\dagger_{s_{2}}~ 
\Big [ 
- 2 {q^\perp \over q^+} + \sigma^\perp {\sigma^\perp.p_1^\perp \over p_1^+}
+ {\sigma^\perp. p_2^\perp \over p_2^+} \sigma^\perp 
 \Big ] ~
\chi_{s_{1}}~. ~(\epsilon^\perp_\lambda)^* 
\nonumber \\
&& ~~~~~~~{g^2 \over 8 \pi^2}  ~ {1 \over 2}~ C_A 
~ ln{\Lambda \over \mu} 
~2 ~ln{p_1^+ p_2^+ \over q^+
\epsilon}.\label{eq3a1}
\end{eqnarray}
The boost invariance of this result is again clear.
Divergent contributions that contain single logarithm are
\begin{eqnarray}
{\cal V}_{3a}^{II}~= &&~g ~T^a ~  \sqrt{p_1^+~p_2^+} ~\chi^\dagger_{s_{2}}~ 
\Big [ 6 {q^\perp \over q^+} - 6 {p_1^\perp \over p_1^+} + {\sigma^\perp.
p_2^\perp \over p_1^+} \sigma^\perp - {p_2^+ \over p_1^+} {\sigma^\perp .
p_1^\perp \over p_1^+} \sigma^\perp
 \Big ] ~
\chi_{s_{1}}~. ~(\epsilon^\perp_\lambda)^* \nonumber \\
&& ~~~~~~~~~~ {g^2 \over 8 \pi^2}~{1 \over 2}~ C_A~ ln{\Lambda \over \mu}.
\label{eq3a2}
\end{eqnarray}
Expressing the terms inside the square bracket in terms of the internal
momenta we get $ - { 1 \over P^+} \Big [ {6 \kappa_1^\perp \over 1-x} -
\sigma^\perp. \kappa_1^\perp \sigma^\perp \Big ]$ which makes boost invariance
explicit.

The contribution from Fig. 3(b) is
\begin{eqnarray}
{\cal V}_{3b}~=&&~ {g^3 \over 2 (2 \pi)^3} (-i f^{abc}T^b T^c)~
\sqrt{p_1^+~p_2^+} ~
\int_\epsilon^{q^+ - \epsilon} dk_1^+~
\int ~d^2 k_1^\perp ~ \theta(\Lambda - \mid k^\perp \mid )
{ 1 \over k_1^+} ~{1 \over k_2^+}~ { 1 \over p_1^- - k_1^- - k^-}~
\nonumber \\
&& { 1 \over
p_1^- - k_1^- - k_2^- - p_2^-}~ \sum_{\sigma_1 \lambda_1 \lambda_2} 
\chi^\dagger_{s_2} ~ \Big [ -2 { k_2^\perp \over k_2^+} + \sigma^\perp
{\sigma^\perp . k^\perp \over k^+} + {\sigma^\perp. p_2^\perp \over p_2^+}
\sigma^\perp \Big ]~ \chi_{\sigma_1} ~. ~(\epsilon^\perp_{\lambda_{2}})^*~
\nonumber \\
&& ~~~~~ 
\chi^\dagger_{\sigma_1} ~\Big [ -2 { k_1^\perp \over k_1^+} + \sigma^\perp
{\sigma^\perp . p_1^\perp \over p_1^+} + {\sigma^\perp. k^\perp \over k^+}
\sigma^\perp \Big ] ~\chi_{s_{1}} ~.~ (\epsilon^\perp_{\lambda_{1}})^* \nonumber \\
&& ~~~\epsilon^j_{\lambda_{1}} ~(\epsilon^i_{\lambda})^*~
\epsilon^l_{\lambda_{2}}~ \Bigg [ \Big [ (k_1^i - k_2^i) - {q^i \over
q^+} (k_1^+ - k_2^+) \Big ] \delta_{lj} 
- \Big [ (k_1^l + q^l) - {k_2^l \over k_2^+} (k_1^+ + q^+) \Big ] \delta_{ij}
\nonumber \\
&& ~~~~~~~+  \Big [ (q^j + k_2^j) - {k_1^j \over k_1^+} (q^+ + k_2^+) \Big ]
\delta_{il} \Bigg ] .  \label{eq3b}
\end{eqnarray}
Here $ k^{(+,\perp)} = p_1^{(+,\perp)} - k_1^{(+,\perp)}$ and 
$ k_2^{(+,\perp)} =
q^{(+,\perp)} - k_1^{(+,\perp)}$.

Divergent contributions that contain products of logarithms are
\begin{eqnarray}
{\cal V}_{3b}^{I}~=&&~g ~ T^a ~
  \sqrt{p_1^+~p_2^+} ~\chi^\dagger_{s_{2}}~ 
\Big [ 
- 2 {q^\perp \over q^+} + \sigma^\perp {\sigma^\perp.p_1^\perp \over p_1^+}
+ {\sigma^\perp. p_2^\perp \over p_2^+} \sigma^\perp 
 \Big ] ~
\chi_{s_{1}}~. ~(\epsilon^\perp_\lambda)^* ~ \nonumber \\
&& ~~~~~~{g^2 \over  8 \pi^2} ~{1 \over 2} C_A ~   ln{\Lambda \over \mu} ~ 
6~ ln{q^+ \over \epsilon} \label{eq3b1}
\end{eqnarray}
which is manifestly boost invariant.
Divergent contributions that contain single logarithm are
\begin{eqnarray}
{\cal V}_{3b}^{II}~=&&~ g ~ T^a~
   \sqrt{p_1^+}~
\sqrt{p_2^+}~ \chi^\dagger_{s_{2}}~ 
\Big [ - 3 \sigma^\perp {\sigma^\perp. p_1^\perp \over p_1^+} - 3
{\sigma^\perp. p_2^\perp \over p_2^+}\sigma^\perp \nonumber \\
&& ~~~+ 6 {p_1^\perp \over p_1^+}
+ {p_2^+ \over p_1^+}~ {\sigma^\perp. p_1^\perp \over p_1^+} \sigma^\perp
 - {\sigma^\perp.
p_2^\perp \over p_1^+} \sigma^\perp  
 \Big ] ~
\chi_{s_{1}}~. ~(\epsilon^\perp_\lambda)^* 
 {g^2 \over 8 \pi^2}~ {1 \over 2}~ C_A ~ ln{\Lambda \over \mu}.
\label{eq3b2}
\end{eqnarray}
Comparison with eq.(\ref{eq2b2}) again makes the boost invariance of this
answer explicit.
\section{Coupling Constant Renormalization}
For the sake of 
completeness, we present here the results for the other diagrams which are
relevant for the coupling constant renormalization. We also calculate the 
$\beta$-function which exactly matches
with the well known results and therefore extends the results arrived at in
the Ref.(\cite{zh93c}), to the {\it most general kinematics} in the
two-component formalism.

The sum of divergent contributions from Figs. 2(a) and 2(b) is
\begin{eqnarray}
{\cal V}_2~=&&~g T^a~
\sqrt{p_1^+~p_2^+}~ \chi^\dagger_{s_{2}}~ \Big [
-2 {q^\perp \over q^+} + \sigma^\perp {\sigma^\perp. p_1^\perp \over p_1^+}
+ {\sigma^\perp .p_2^\perp \over p_2^+} \sigma^\perp \Big ]~ \chi_{s_{1}}
~.~ (\epsilon^\perp_\lambda)^* ~ \nonumber \\
&& {g^2 \over 8 \pi^2} ~\Big ( -{1 \over 2}~C_A ~+~ C_f \Big )~
ln{\Lambda \over \mu} ~\Big( 2 ~ln{p_1^+ p_2^+ \over \epsilon^2} - 3 \Big ),
\end{eqnarray}
where we observe the emergence of the canonical vertex structure.

The sum of divergent contributions from Figs. 3(a) and 3(b) is
\begin{eqnarray}
{\cal V}_3 ~=&&~ g~ T^a~
 ~ \sqrt{p_1^+}~
\sqrt{p_2^+}~ \chi^\dagger_{s_{2}}~ 
\Big [ 
- 2 {q^\perp \over q^+} + \sigma^\perp {\sigma^\perp.p_1^\perp \over p_1^+}
+ {\sigma^\perp. p_2^\perp \over p_2^+} \sigma^\perp 
 \Big ] ~
\chi_{s_{1}}~.~ (\epsilon^\perp_\lambda)^*  \nonumber \\
&& ~~~~~~ {g^2 \over 8 \pi^2} {1 \over 2}~ C_A ~
ln{\Lambda \over \mu}
~ \Big ( 2~ ln{p_1^+ p_2^+ \over \epsilon^2}~ +~ 4~ ln{q^+ \over \epsilon}
~ -~ 3
\Big) ,
\end{eqnarray}
where we again observe the emergence of the canonical vertex.

The diagrams in Figs. 4(a), 4(b), 5(a) and 5(b) correspond to the renormalization of the
external quark and gluon legs that are connected to the vertex. Their
contributions are given below.
\begin{eqnarray}
{\cal V}_{4a}=g~ T^a~
 ~ \sqrt{p_1^+ p_2^+}~ \chi^\dagger_{s_{2}}~ 
&& ~~\Big [ 
- 2 {q^\perp \over q^+} + \sigma^\perp {\sigma^\perp.p_1^\perp \over p_1^+}
+ {\sigma^\perp. p_2^\perp \over p_2^+} \sigma^\perp 
 \Big ] ~
\chi_{s_{1}}~.~ (\epsilon^\perp_\lambda)^*  \nonumber \\
&& ~~~~~~ {g^2 \over 4 \pi^2} ~ C_f~
ln{\Lambda \over \mu}
~ \Big ( {3 \over 2}- 2~ln{p_1^+  \over \epsilon}~ 
\Big) ,
\end{eqnarray}
\begin{eqnarray}
{\cal V}_{4b}=g~ T^a~
 ~ \sqrt{p_1^+ p_2^+}~ \chi^\dagger_{s_{2}}~ 
&& ~~\Big [ 
- 2 {q^\perp \over q^+} + \sigma^\perp {\sigma^\perp.p_1^\perp \over p_1^+}
+ {\sigma^\perp. p_2^\perp \over p_2^+} \sigma^\perp 
 \Big ] ~
\chi_{s_{1}}~.~ (\epsilon^\perp_\lambda)^*  \nonumber \\
&& ~~~~~~ {g^2 \over 4 \pi^2} ~ C_f~
ln{\Lambda \over \mu}
~ \Big ( {3 \over 2}- 2~ln{ p_2^+ \over \epsilon}~ 
\Big) ,
\end{eqnarray}
\begin{eqnarray}
{\cal V}_{5a}=-g~ T^a~
 ~ \sqrt{p_1^+ p_2^+}~ \chi^\dagger_{s_{2}}~ 
&& ~~\Big [ 
- 2 {q^\perp \over q^+} + \sigma^\perp {\sigma^\perp.p_1^\perp \over p_1^+}
+ {\sigma^\perp. p_2^\perp \over p_2^+} \sigma^\perp 
 \Big ] ~
\chi_{s_{1}}~.~ (\epsilon^\perp_\lambda)^*  \nonumber \\
&& ~~~~~~ {g^2 \over 8 \pi^2} {4 \over 3}~ N_fT_f~
ln{\Lambda \over \mu} .
\end{eqnarray}
\begin{eqnarray}
{\cal V}_{5b}=g~ T^a~
 ~ \sqrt{p_1^+ p_2^+}~ \chi^\dagger_{s_{2}}~ 
&& ~~\Big [ 
- 2 {q^\perp \over q^+} + \sigma^\perp {\sigma^\perp.p_1^\perp \over p_1^+}
+ {\sigma^\perp. p_2^\perp \over p_2^+} \sigma^\perp 
 \Big ] ~
\chi_{s_{1}}~.~ (\epsilon^\perp_\lambda)^*  \nonumber \\
&& ~~~~~~ {g^2 \over 8 \pi^2} ~ C_A~
ln{\Lambda \over \mu}
~ \Big ( {11 \over 3}- 4~ln{q^+  \over \epsilon}~ 
\Big ) ,
\end{eqnarray}

Now, to evaluate the contributions to the coupling constant, we have to
multiply ${\cal V}_4$ and ${\cal V}_5$ with ${1\over2}$ in order to take into account
the proper correction due to the renormalization of initial and final states
\cite{bd}. Thus adding the contributions we get,
\begin{eqnarray}
\delta {\cal V}_1 =&&~~\big( {1\over2}{\cal V}_4 + {1\over 2}{\cal V}_5 +
{\cal V}_2 +
{\cal V}_3 \big)
\nonumber\\=&&~~g~ T^a~
 ~ \sqrt{p_1^+ p_2^+}~ \chi^\dagger_{s_{2}}~ 
\Big [ 
- 2 {q^\perp \over q^+} + \sigma^\perp {\sigma^\perp.p_1^\perp \over p_1^+}
+ {\sigma^\perp. p_2^\perp \over p_2^+} \sigma^\perp 
 \Big ] ~
\chi_{s_{1}}~.~ (\epsilon^\perp_\lambda)^*  \nonumber \\&&\quad\quad
{g^2\over 8\pi^2}\Big( {11\over6} C_A-{2 \over 3}N_fT_f\Big)~ln {\Lambda\over
\mu}
\end{eqnarray}
Note that all the mixed divergences cancel. The correction to the coupling
constant is given by
\begin{eqnarray}
g_R~=~g(1+\delta g)~=~g\Big[~1+ {g^2\over 8\pi^2}\Big( {11\over6} C_A-{2 \over 
3}N_fT_f\Big)ln{\Lambda\over \mu}~\Big].
\end{eqnarray}
We compute the $\beta$-function as 
\begin{eqnarray}
\beta(g)~&&=~-~{\partial{g_R} \over \partial{ln\Lambda}}\nonumber\\
&&=~-~{g^3 \over 16\pi^2}\Big({11\over 3}C_A-{4 \over 
3}N_fT_f\Big) ,
\end{eqnarray}
which is well known result to the one-loop order.
\section{Summary, Discussion and Conclusions}
Calculations employing time ($x^+$) ordered perturbative techniques in
light-front theory are known to be straightforward but long and tedious. A
lot of effort has to be invested in the calculation of coefficient functions
accompanying the divergences for individual diagrams.
To the best of our knowledge, this is the {\it first work} to investigate the
utility of Galilean boost symmetry in determining the correctness of the 
structure of the coefficient functions accompanying the
divergences in light-front perturbation theory beyond tree level.

In this initial investigation we have employed the simplest choice of regulators 
that cutoff single particle momenta.  
One should note that in addition to possible violations of boost invariance,
such simple minded cutoff procedure could in principle even introduce
non-analyticities in the structure of counterterms (see Sec. VI of Ref.
\cite{wi94} for an explicit example). However, in the case of vertex
diagrams, we encounter only logarithmic transverse divergences.
Even with finite cutoffs,
violations of transverse boost invariance can appear only inside the 
logarithms and we
expect the symmetry to be 
present in the non-trivial structure of the coefficient functions 
that accompany the 
divergences. We are primarily interested in understanding the complex
structure of these coefficient functions on the basis of Galilean symmetry.
Incidentally we note that, in contrast, longitudinal boost invariance is a scale 
invariance in
light-front theory. 
The implication of longitudinal boost symmetry for the
coefficient functions is trivial, namely, simple scaling behavior.

Let us summarize our findings. Out of all the  $x^+$-ordered diagrams
relevant for our calculation, four involve wavefunction renormalization
correction and have the structure of the canonical vertex. For the remaining
diagrams which correspond to vertex corrections, 
the divergent contributions from each of the
them contain terms that involve (I) product of logarithms 
and (II) single
logarithm. For contributions that belong to (I), we find that 
for each diagram separately, 
the coefficient of the divergent factor is proportional to the canonical
vertex and hence Galilean boost invariance is manifestly maintained. For
contributions that belong to (II), for each diagram, the coefficient of the
divergent factor is not proportional to the canonical vertex. Nevertheless,
in each case, rewriting the coefficient in terms of the internal momenta 
explicitly shows that the coefficient is independent of the total transverse
momentum $P^\perp$. Hence for the contributions that belong to (II) the
constraint from transverse boost invariance is maintained, even though the
canonical form is not reproduced. 

Our results show that two-dimensional Galilean invariance which is
manifest at tree level is also exhibited in the coefficient functions
accompanying the divergences in the regulated theory at the one loop level in
the case of quark-gluon vertex in light-front QCD even with a regulator that
violates the symmetry. Since the symmetry is only a part of the complete
Lorentz symmetry, we expect the constraints which follow from the
invariance to be less restrictive. Indeed, our results show
that the structure of
the vertex that satisfies transverse boost invariance is not unique.

Even-though the canonical vertex structure is not reproduced in the
coefficient of the single logarithms, it still has some usefulness in
practical calculations since it obeys constraint from Galilean boost 
invariance. The coefficient functions accompanying single logarithms are
obtained after isolating the leading double logarithms and they exhibit a
complicated structure. It is quite easy to
make a mistake in the sign  in {\it one} of the terms for individual $x^+$
ordered diagrams. Our calculations show that using the underlying
transverse boost
symmetry one can easily recognize the mistake in the calculation and hence
correct it.

We have also summarized the results for the
complete set of diagrams contributing to coupling constant renormalization
for the massless quark case. We have extracted the $ \beta $-function which
matches with the well-known results and therefore extends the results
arrived previously to the most general kinematics.
%
%
Using the
two-component representation \cite{zh93b} 
we have presented {\it for the
first time} the
results separately for each $x^+$ ordered diagram with arbitrary 
external momenta which is essential to study the renormalization of the
helicity-non flip parts of the vertex. Present calculations together with the
calculations presented in Ref. \cite{zh93c} explicitly show that linear
divergences of the type ${ 1 \over \epsilon}$ where $\epsilon$ is the cutoff
on longitudinal loop momenta occur in individual time-ordered diagrams only
in radiative corrections to the chiral symmetry breaking part of the
quark gluon vertex. This divergence is a special feature of non-abelian
gauge theory. At one loop level, this divergence cancels with our
choice of regulators when
different time-ordered diagrams are summed up. Since intermediate states
involved are, in general, different in different time ordered diagrams, the
cancellation may no longer be operative once more sophisticated regulators
that explicitly depend on the intermediate states are employed. This needs
to be investigated in detail in the future because of its nontrivial
consequences for the renormalization of chiral symmetry breaking terms in the 
QCD Hamiltonian. 

The present calculations are also essential for the development of a new
method \cite{deep} of calculation of structure functions in deep inelastic
scattering. This approach combines the techniques of light-front current
algebra and Fock space expansion for the Hamiltonian in the light-front
gauge $A^+=0$, to treat the non-perturbative and perturbative parts of the
structure functions in the same language, namely, that of multi-parton
wavefunctions. Up to now, renormalization has been performed in this
framework to second order for unpolarized and polarized structure functions
in perturbative QCD \cite{pp}. The calculations presented in this paper constitute 
essential parts of a complete fourth order analysis of leading logarithms
which is necessary to establish the viability of the new approach.

\appendix
\section{Manifest boost symmetry of energy denominators and vertices}
In this appendix we verify the Galilean boost invariance of vertices and
energy differences that occur in light-front time-ordered loop diagrams. 
First consider the canonical vertex given in eq.(\ref{canon}). 
Let $P^+$ and $P^\perp$  denote total longitudinal and transverse momentum in
the problem. We introduce the momentum fractions $x_i$ and the relative
transverse momenta $\kappa_i^\perp$ by 
\begin{eqnarray}
 p_2^+ = x P^+~,~ p_2^\perp =
\kappa_1^\perp + x P^\perp, ~~ q^+ = (1-x)P^+~,~ q^\perp = - \kappa_1^\perp +
(1-x) P^\perp.
\end{eqnarray}
The longitudinal momentum fractions $x_i$ and the relative
transverse momenta $\kappa_i^\perp$ obey the constraints $\sum x_i =1$ and
$\sum \kappa_i^\perp = 0$. The canonical vertex takes the form
\begin{eqnarray}
{\cal V}_0 ~= ~g ~T^a ~ \sqrt{x}~ \chi^\dagger_{s_{2}}~
 \Big [  2 {\kappa_1^\perp \over 1-x}
+ {\sigma^\perp
.\kappa_1^\perp \over x} \sigma^\perp  +im \Big( 1  - {1 \over
x} \Big) \Big ] ~\chi_{s_{1}} ~ .~
(\epsilon^\perp_\lambda)^*. 
\end{eqnarray}
In terms of the internal momenta, the boost invariance of the quark-gluon
vertex is clearly manifest.

Next consider loop diagrams. 
As an example we consider the diagram shown in Fig. 2(a).
Parameterize the single particle momenta in terms of the internal momenta 
as follows. 
\begin{eqnarray}
k_3^+ = y P^+~,~ k_3^\perp = \kappa_2^\perp + y P^\perp, ~~
k_1^+=(1-y)P^+~,~ k_1^\perp = - \kappa_2^\perp + (1-y) P^\perp.
\end{eqnarray}
Then 
\begin{eqnarray}
k^+ = k_1^+ - q^+ = (x-y) P^+, ~~ k^\perp = k_1^\perp - q^\perp =
\kappa_1^\perp - \kappa_2^\perp  + (x-y) P^\perp.
\end{eqnarray}
The energy difference appearing in the two energy denominators are, then,
\begin{eqnarray} 
p_1^- - k_1^- - k_3^- = &&- {(\kappa_2^\perp)^2 \over P^+} \Big ( {1 \over y} +
{1 \over 1-y } \Big ), \nonumber \\
~~ p_1^- - k_3^- - k^- - q^- = && - { 1 \over P^+} \Big
[ {(\kappa_2^\perp)^2 \over y} + {(\kappa_1^\perp)^2 \over 1-x} +
{(\kappa_1^\perp - \kappa_2^\perp)^2 \over x-y} \Big ].
\end{eqnarray}
The vertex factors are 
\begin{eqnarray}
 -2  {k_3^\perp \over k_3^+} + \sigma^\perp {\sigma^\perp .k^\perp \over
k^+} + {\sigma^\perp . p_2^\perp \over p_2^+} \sigma^\perp
= &&{1 \over P^+} \Big [- 2 {(\kappa_1^\perp - \kappa_2^\perp) \over x-y} -
\sigma^\perp . {\sigma^\perp . \kappa_2^\perp \over 1-y} + {\sigma^\perp .
\kappa_1^\perp \over x} \sigma^\perp \Big ], 
\nonumber \\
 -2  {q^\perp \over q^+} + \sigma^\perp {\sigma^\perp .k_1^\perp \over
k_1^+} + {\sigma^\perp . k^\perp \over k^+} \sigma^\perp 
= &&{1 \over P^+} \Big [ 2 {\kappa_1^\perp \over 1-x} + \sigma^\perp 
{\sigma^\perp . \kappa_2^\perp \over y} - {\sigma^\perp . \kappa_2^\perp
\over 1-y} \sigma^\perp \Big ], \nonumber \\
-2  {k_3^\perp \over k_3^+} + \sigma^\perp {\sigma^\perp .p_1^\perp 
\over p_1^+} + {\sigma^\perp . k_1^\perp \over k_1^+} \sigma^\perp
= &&{ 1 \over P^+} \Big [ - 2 {(\kappa_1^\perp - \kappa_2^\perp) \over x -y} +
{\sigma^\perp . \kappa_2^\perp \over y} \sigma^\perp \Big ].
\end{eqnarray}
Thus the vertices and energy denominators appearing in Fig. 2(a) are
manifestly invariant under the Galilean boosts in the transverse plane and
this is a general property of any $x^+$ ordered diagram in light-front
perturbation theory.


\eject         
\vskip .5in
\centerline{\bf List of Figures}
\vskip .5in
\begin{enumerate}
\item The canonical quark-gluon vertex in light-front QCD.
\item Contribution to quark-gluon vertex from contributions involving two
quark-gluon vertices.
\item Contribution to quark-gluon vertex from contributions involving one
quark-gluon vertex and one three gluon vertex.
\item Contribution to quark-gluon vertex from fermion wavefunction
renormalization.
\item Contribution to quark-gluon vertex from gluon wavefunction
renormalization.
\end{enumerate}
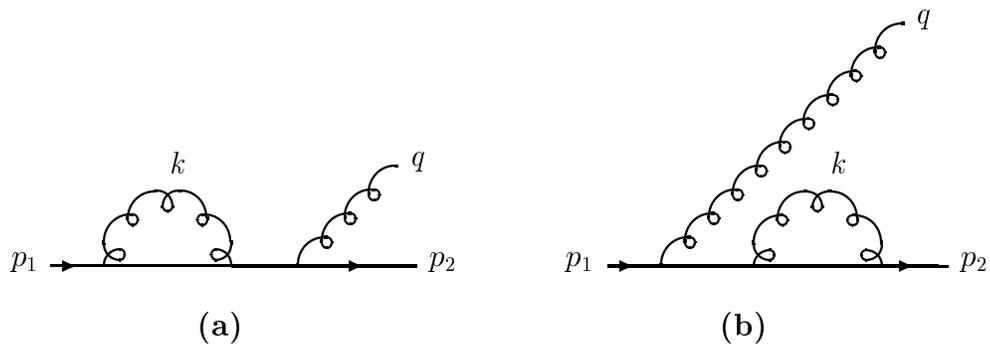
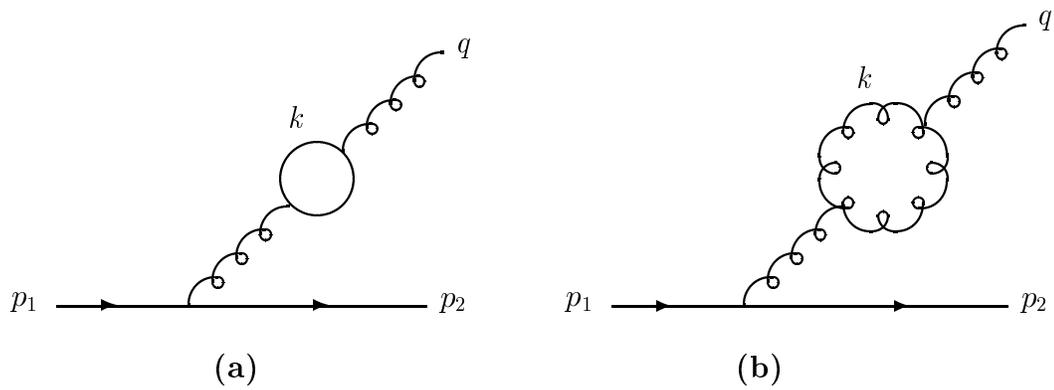
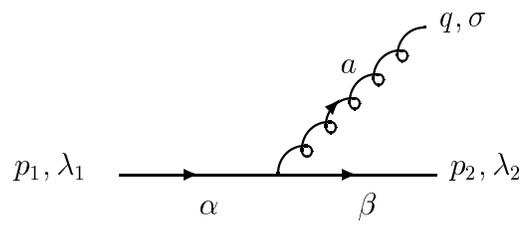
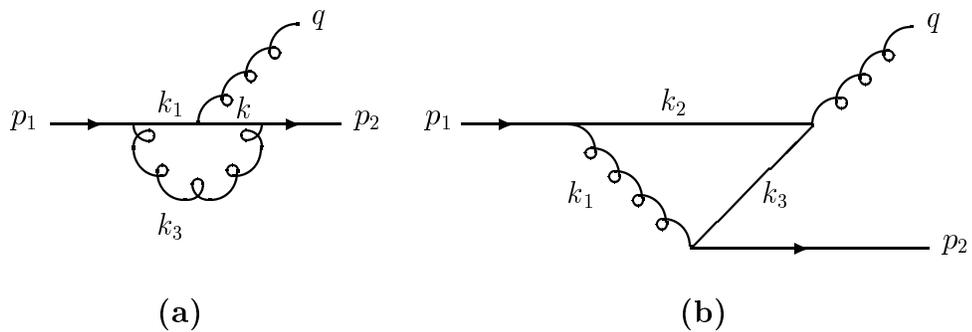
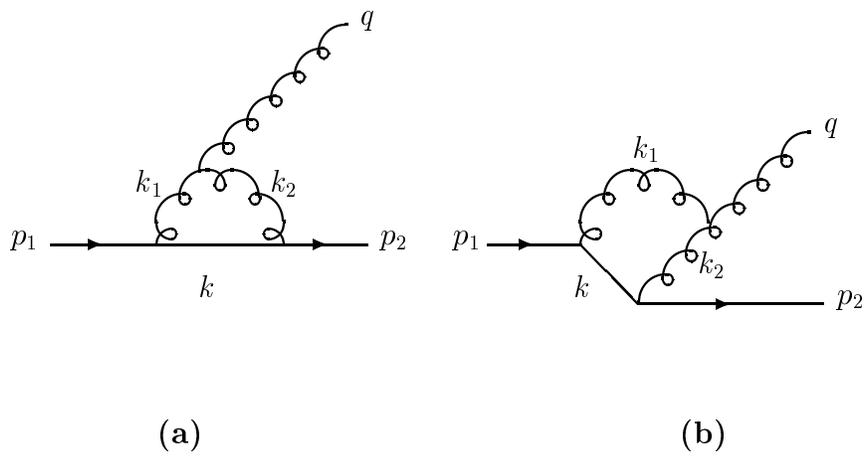
\end{document}